\documentclass[sigconf,screen]{acmart}

\AtBeginDocument{%
  \providecommand\BibTeX{{%
    \normalfont B\kern-0.5em{\scshape i\kern-0.25em b}\kern-0.8em\TeX}}}

\usepackage{amsfonts}
\usepackage{subcaption}
\usepackage[ruled,linesnumbered]{algorithm2e}
\usepackage{enumitem}
\usepackage{color}
\usepackage{url}
\usepackage{bm}
\usepackage{bbold}
\usepackage{multirow}

\DeclareMathOperator*{\argmax}{\arg\max}
\usepackage{makecell}
\usepackage{amsthm}

\newtheoremstyle{bolddefinition} 
{}          
{}          
{}          
{}          
{\bfseries} 
{.}         
{.5em}         
{\thmname{#1}\thmnumber{ #2}\thmnote{ (\bfseries #3)}}          

\theoremstyle{bolddefinition}
\newtheorem{definition}{Definition}

\usepackage{listings}

\definecolor{backgroundcolor}{RGB}{242, 242, 235}
\definecolor{keywordcolor}{RGB}{67, 65, 249}
\definecolor{stringcolor}{RGB}{169, 66, 221}

\definecolor{lightyellow}{RGB}{255, 247, 224}

\newcounter{listing}

\newcommand{\mypara}[1]{\smallskip \noindent\textbf{#1.} \xspace}

\usepackage[most]{tcolorbox}
\tcbuselibrary{skins, breakable, theorems}
\usepackage{colortbl}

\newcommand{\system}{\textsc{CodeEraser \xspace}}%
\newcommand{\systemnospace}{\textsc{CodeEraser}}%

\definecolor{lightcyan}{RGB}{243,248,252}

\usepackage{framed}
\newenvironment{summary}{%
  \def\FrameCommand##1{%
    \setlength{\fboxrule}{1pt}%
    \setlength{\fboxsep}{4.5pt}%
    \fcolorbox{black}{lightcyan}{##1}%
  }%
  \MakeFramed{\advance\hsize-\width \FrameRestore}%
  \setlength{\parindent}{0pt}
  \noindent\ignorespaces
}{%
  \endMakeFramed
}

\definecolor{matchgreen}{RGB}{224,243,220}
\newcommand{\revise}[1]{{\textcolor{black}{#1}}}

\copyrightyear{2026}
\acmYear{2026}
\setcopyright{rightsretained}
\acmConference[ICSE '26]{2026 IEEE/ACM 48th International Conference on Software Engineering}{April 12--18, 2026}{Rio de Janeiro, Brazil}
\acmBooktitle{2026 IEEE/ACM 48th International Conference on Software Engineering (ICSE '26), April 12--18, 2026, Rio de Janeiro, Brazil}
\acmPrice{}
\acmDOI{10.1145/3744916.3764573}
\acmISBN{979-8-4007-2025-3/26/04}

\author{Zhaoyang Chu}
\authornote{Also with National Engineering Research Center for Big Data Technology and System, Services Computing Technology and System Lab, Cluster and Grid Computing Lab, School of Computer Science and Technology, Huazhong University of Science and Technology, Wuhan, 430074, China.}
\affiliation{%
 \institution{Huazhong University of Science and Technology}
 \country{Wuhan, China}
 }
\email{chuzhaoyang@hust.edu.cn}

\author{Yao Wan}
\authornotemark[1]
\authornote{Yao Wan is the corresponding author.}
\affiliation{%
 \institution{Huazhong University of Science and Technology}
 \country{Wuhan, China}
 }
\email{wanyao@hust.edu.cn}

\author{Zhikun Zhang}
\affiliation{%
\institution{Zhejiang University}
\country{Hangzhou, China}
}
\email{zhikun@zju.edu.cn}

\author{Di Wang}
\affiliation{%
\institution{King Abdullah University of Science and Technology}
\country{Thuwal, Saudi Arabia}
}
\email{di.wang@kaust.edu.sa}

\author{Zhou Yang}
\affiliation{%
\institution{University of Alberta}
\country{Edmonton, Canada}
}
\email{zy25@ualberta.ca}

\author{Hongyu Zhang}
\affiliation{%
\institution{Chongqing University}
\country{Chongqing, China}
}
\email{hyzhang@cqu.edu.cn}

\author{Pan Zhou}
\affiliation{%
 \institution{Huazhong University of Science and Technology}
 \country{Wuhan, China}
 }
\email{panzhou@hust.edu.cn}

\author{Xuanhua Shi}
\authornotemark[1]
\affiliation{%
 \institution{Huazhong University of Science and Technology}
 \country{Wuhan, China}
 }
\email{xhshi@hust.edu.cn}

\author{Hai Jin}
\authornotemark[1]
\affiliation{%
 \institution{Huazhong University of Science and Technology}
 \country{Wuhan, China}
 }
\email{hjin@hust.edu.cn}

\author{David Lo}
\affiliation{%
\institution{Singapore Management University}
\country{Singapore, Singapore}
}
\email{davidlo@smu.edu.sg}

\begin{document}

\title{Scrub It Out! Erasing Sensitive Memorization in Code Language Models via Machine Unlearning}

\begin{abstract}
While \textit{Code Language Models} (CLMs) have demonstrated superior performance in software engineering tasks such as code generation and summarization, recent empirical studies reveal a critical privacy vulnerability: these models exhibit unintended memorization of sensitive training data, enabling verbatim reproduction of confidential information when specifically prompted. 
To address this issue, several approaches, including training data de-duplication and differential privacy augmentation, have been proposed. However, these methods require full-model retraining for deployed CLMs, which incurs substantial computational costs.
In this paper, we aim to answer the following research question: 
\textit{Can sensitive information memorized by CLMs be erased effectively and efficiently?}

We conduct a pioneering investigation into erasing sensitive memorization in CLMs through machine unlearning—a \textit{post-hoc} modification method that removes specific information from trained models without requiring full retraining. 
Specifically, we first quantify the memorization risks of sensitive data within CLM training datasets and 
\revise{curate a high-risk dataset of 50,000 sensitive memorized samples as unlearning targets.}
\revise{We study two widely used gradient ascent-based unlearning approaches: the vanilla and constraint-based methods,}
\revise{and introduce \systemnospace, an advanced variant that}
\textit{selectively} unlearns sensitive memorized segments in code while preserving the structural integrity and functional correctness of the surrounding code.
Extensive experiments on three families of CLMs, \textit{i.e.}, CodeParrot, CodeGen-Mono, and Qwen2.5-Coder, validate the effectiveness and efficiency of \system in erasing targeted sensitive memorization while maintaining model utility.
\end{abstract}



\maketitle

\section{Introduction}
\label{sec_intro}

Recently, \textit{Code Language Models} (CLMs), such as CodeGen~\cite{Nijkamp2023codegen}, Code Llama~\cite{Roziere2023codellama}, and Qwen2.5-Coder~\cite{hui2024qwen25coder}, have demonstrated significant potential in automating various aspects of software engineering, including code generation~\cite{Du2023classeval, Jiang2023self_plan_code_gen, wang2025codesync}, code summarization~\cite{AlKaswan2023extending_to_summarize, Arakelyan2023exploring_llms_code_analysis, wan18rl_code_sum}, program repair~\cite{Xia2023automated_prog_repair, Xia2023conversational_prog_repair}, and type inference~\cite{li2025data_flow_guided}. 
Their success can be attributed to pre-training with autoregressive language modeling on large-scale code corpora~\cite{wan2022what_capture, wan2024dl_code_survey}, where the model predicts the next token given a sequence of previous tokens.

\begin{figure*}[!t]
	\centering
	\includegraphics[width=0.98\textwidth]{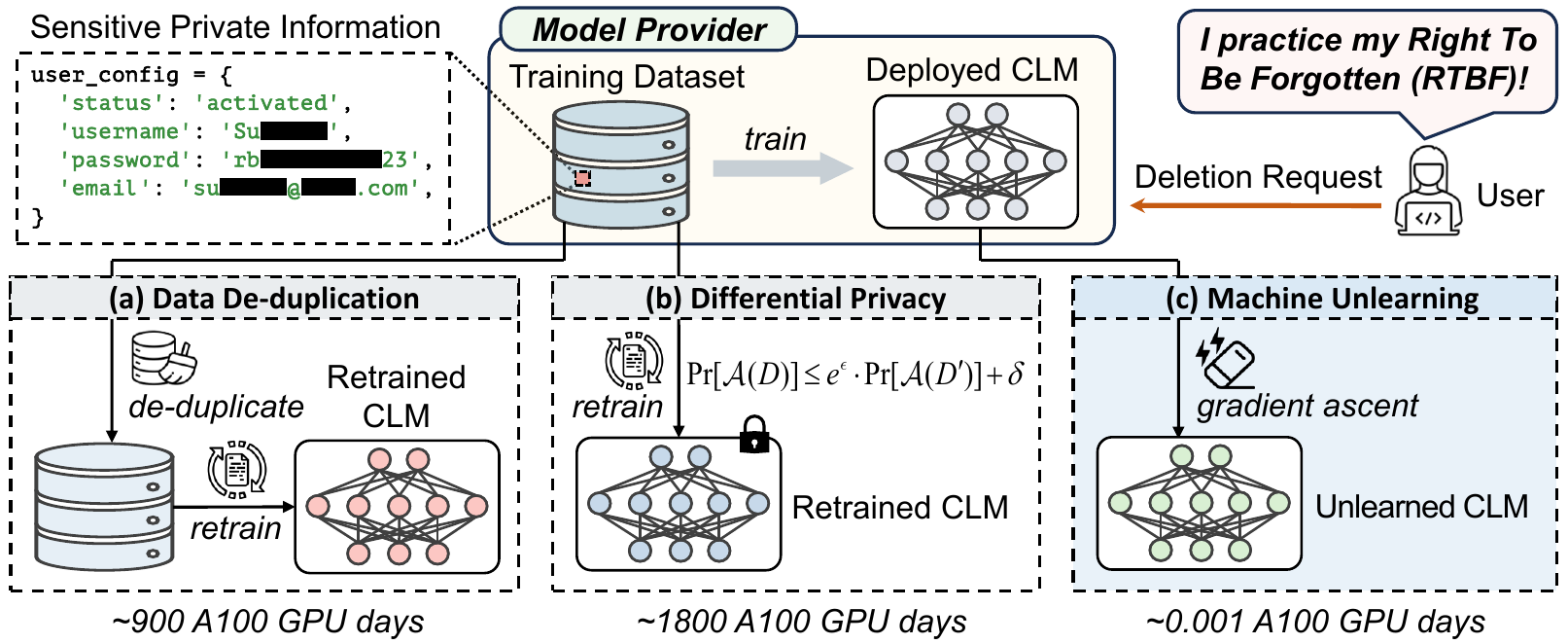}
         \vspace{-0.5em}
	\caption{An illustration of existing methods, \textit{i.e.}, (a) data de-duplication, (b) differential privacy, and (c) machine unlearning, to mitigate the memorization of sensitive information in CLMs. We mask the personal details for ethical considerations.}
	\label{fig_illustration}
	\vspace{-0.5em}
\end{figure*}

However, studies have shown that the current pre-training paradigm may retain sensitive data, \textit{e.g.}, emails and passwords, encountered during the training phase~\cite{AlKaswan2024code_memorization_traces, Carlini2021extracting, Carlini2023quantifying_memorization, Niu2023codexleaks, Yang2024code_model_memorization}. 
This retention occurs because CLMs, trained on vast amounts of code collected from GitHub repositories, may inadvertently memorize sensitive data embedded within these repositories, including personally identifiable information (\textit{e.g.}, \textit{names}, \textit{emails}, and \textit{phone numbers}) and private authentication credentials (\textit{e.g.}, \textit{passwords}, \textit{API keys}, and \textit{cryptographic secrets})~\cite{Basak2023secretbench, Feng2022secret_leakage_detection, Huang2024code_secret_memorization, Meli2019github_secret_leakage, Niu2023codexleaks, Yang2024code_model_memorization}.

From another perspective, to strengthen individual control over personal data, global legislative frameworks such as the European Union's \textit{General Data Protection Regulation} (GDPR)~\cite{European2018gdpr} and the \textit{California Consumer Privacy Act} (CCPA)~\cite{California2023ccpa} have established the ``\textit{Right to Be Forgotten}'' (RTBF)~\cite{Veale2018algorithms_remember, Villaronga2018rtbf}. These regulations empower individuals to request the deletion of their personal data, providing a critical safeguard for privacy protection.

To mitigate the privacy risks posed by potential data leaks in CLMs and ensure compliance with the RTBF, we investigate the following question: \textit{Can sensitive information memorized by CLMs be erased effectively and efficiently?}

\mypara{Intuitive Approaches and Limitations}
Our investigation reveals two distinct research directions. The first focuses on training data de-duplication, as illustrated in~\autoref{fig_illustration} (a). Prior studies~\cite{Carlini2023quantifying_memorization, Kandpal2022deduplicating_for_privacy, Lee2022deduplicating} have demonstrated that de-duplication can mitigate the memorization tendencies of LMs. However, experimental evidence indicates that LMs still retain substantial memorization capabilities even under this paradigm~\cite{Biderman2023predictable_memorization, Biderman2023pythia, Jang2023knowledge_unlearning}.

Another line of research falls into \textit{Differential Privacy} (DP)~\cite{Abadi2016deep_dp, Zhu2022dp_in_ai}, as shown in~\autoref{fig_illustration} (b).
This approach enforces the formal guarantee that the addition or removal of any training data point does not substantially affect the final model~\cite{Zhu2017dp_survey}, thereby providing formal privacy guarantees for individual training samples. However, DP-based training fundamentally limits LMs' ability to capture long-tail patterns in data distributions, resulting in significant utility degradation~\cite{Anil2022dp_bert, Feldman2020_deep_long_tail, Feldman2020discover_long_tail}.

Furthermore, both DP and de-duplication methods are typically applied during the initial training phase.
For already deployed CLMs, these methods lack the ability to selectively remove specific data as requested by users, often necessitating retraining the entire model~\cite{chen2023efficient_unlearn, Jang2023knowledge_unlearning, Nguyen2022unlearning_survey}.
Such retraining is costly and time-consuming, especially given the escalating scale of contemporary CLMs.
This limitation prevents their flexibility in addressing dynamic user requests and evolving privacy demands in real-world scenarios.

\mypara{Our Work: A Machine Unlearning Perspective}
More recently, machine unlearning has emerged as a promising alternative for LMs, as shown in~\autoref{fig_illustration} (c), which seeks to remove specific information by \textit{post-hoc} modifying the trained model~\cite{Bourtoule2021sisa_unlearning, Cao2015machine_unlearning, chen2023efficient_unlearn, Golatkar2020deep_network_scrubbing, Jang2023knowledge_unlearning}.
Existing approaches typically employ gradient ascent to reverse the learning of specific data, thus proactively removing its influence~\cite{chen2023efficient_unlearn, Jang2023knowledge_unlearning}.
Compared to DP and de-duplication techniques, machine unlearning enables LMs to quickly forget certain information with just a few parameter updates without full retraining, thereby reducing the training time from 900$\sim$1800 to $\sim$0.001 A100 GPU days~\cite{Jang2023knowledge_unlearning}.
Nevertheless, we argue that these approaches often indiscriminately forget entire text instances rather than selectively targeting specific sensitive information. 
As a result, they struggle to erase sensitive segments (\textit{e.g.}, API key strings) embedded in code without disrupting 
the structural integrity and functional correctness of the surrounding code.

In this paper, we perform a pioneering investigation into sensitive memorization erasure
\footnote{\revise{We adopt the term \textit{sensitive memorization} to denote the phenomenon where CLMs retain and reproduce sensitive training data (\textit{e.g.}, API keys). Thus, by \textit{memorization erasure}, we mean techniques designed to remove such retained sensitive content.}}
in CLMs through machine unlearning.
Specifically, we first quantify the memorization risks of sensitive data within CLM training corpora
and curate a high-risk dataset of 50,000 sensitive memorized samples as unlearning targets. 
We investigate two widely used gradient ascent-based unlearning approaches: the vanilla method and the constraint-based method, and further develop an advanced variant, termed \systemnospace, which selectively unlearns sensitive memorized segments in code while preserving 
the surrounding code's integrity and functionality.

To assess the effectiveness and efficiency of \systemnospace, we conduct extensive experiments on three widely used suites of CLMs, \textit{i.e.}, CodeParrot~\cite{HuggingFace2022codeparrot}, CodeGen-Mono~\cite{Nijkamp2023codegen}, and Qwen2.5-Coder~\cite{hui2024qwen25coder}.
The results demonstrate \systemnospace's ability to effectively and efficiently mitigate the memorization issue in CLMs, thus protecting sensitive data against potential extraction attacks.
Using the Qwen2.5-Coder-7B model as an example, \system successfully reduces memorization by 93.89\% on the targeted forgotten set (sampled from the sensitive memorization dataset), while retaining 99.00\% of the model's original performance, 
with an average processing time of 46.88 seconds per sample.

\mypara{Contributions}
The key contributions of this paper are as follows.
\begin{itemize}[leftmargin=4mm, itemsep=0.05mm]

\item 
\textbf{New Problem and Dataset.}
To the best of our knowledge, we are the first to formulate the problem of erasing sensitive memorization within CLMs. As an initial step, we curate a sensitive memorization dataset to support further research in this area.

\item 
\textbf{Pioneering Study.}
We conduct the first comprehensive study on sensitive memorization erasure in CLMs via machine unlearning. 
We also introduce a selective gradient-ascent approach \system to target and remove sensitive memorized segments while preserving code integrity.

\item 
\textbf{Extensive Evaluation.}
We conduct comprehensive experiments on three widely used families of CLMs, namely CodeParrot~\cite{HuggingFace2022codeparrot}, CodeGen-Mono~\cite{Nijkamp2023codegen}, and Qwen2.5-Coder~\cite{hui2024qwen25coder}. The results demonstrate the effectiveness and efficiency of \system in erasing sensitive memorization within CLMs while maintaining acceptable levels of model utility.
\end{itemize}

\section{Background}
\revise{We begin by introducing the background of LMs, followed by a formal definition of memorization in these models.}

\subsection{Language Models}
\label{sec_code_lm}

\textit{Language Models} (LMs) are designed to predict the probability of a token sequence by utilizing the empirical distribution of token occurrences in the training data. One widely adopted unsupervised approach for training LMs is autoregressive language modeling, also known as ``next-token prediction'', where the model sequentially predicts tokens from left to right~\cite{Mikolov2010recurrent, Radford2019gpt2, Brown2020gpt3}.

\mypara{Autoregressive Language Modeling}
Given a token sequence $\mathbf{x} = (x_1, x_2, \ldots, x_N)$, the LM employs the chain rule to model its joint probability as the product of conditional probabilities:
\begin{equation}
    \mathbf{Pr}(x_1, x_2, \ldots, x_N) = \prod_{i=1}^N \mathbf{Pr}(x_i \mid x_1, \ldots, x_{i-1})\,.
\label{equ_language_modeling}
\end{equation}
In this paradigm, a neural network model $f$, parameterized by $\theta$, is typically employed to estimate the likelihood of each token $x_i$ conditioned on preceding tokens, denoted as $f_{\theta}(x_i \mid x_1, \ldots, x_{i-1})$. 
The parameters $\theta$ are optimized by maximizing the probability of each sample within the training dataset $\mathcal{D}$. 
This is achieved by minimizing the loss function as follows:
\begin{equation}
    \mathcal{L}^{LM}(\mathbf{x}) = - \log \prod_{i=1}^N f_{\theta}(x_i \mid x_1, \ldots, x_{i-1}) \,.
\label{equ_loss}
\end{equation}

\begin{table*}[!t]
    \centering
    \caption{A toy example to illustrate the calculation process of MA and $\text{EL}_n$ with $n=3$.}
    \vspace{-0.5em}
    \includegraphics[width=0.98\textwidth]{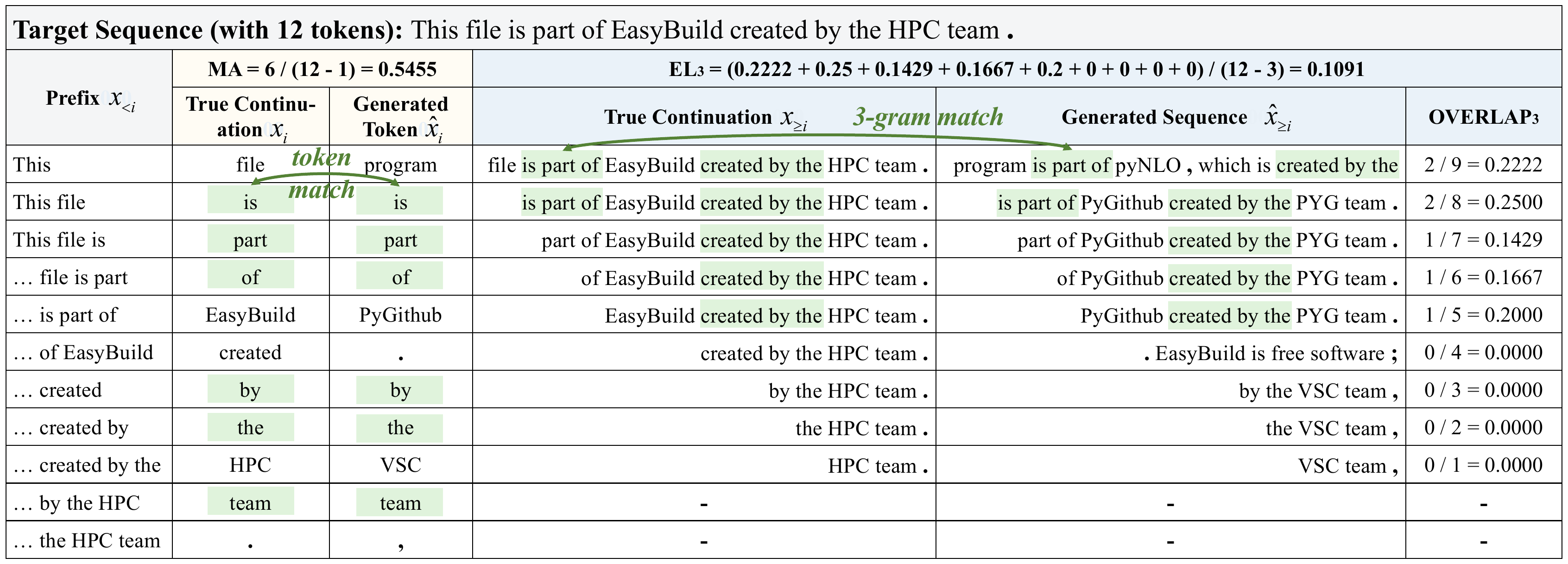}
    \label{tab_memorization_metrics}
    \vspace{-0.5em}
\end{table*}

\mypara{Model Inference via Prefix Prompt}
Once the LM $f_{\theta}$ is trained, it can generate outputs based on a given prefix prompt during inference.
Given a prefix prompt $p = (x_1, \ldots, x_{i-1})$, the trained LM iteratively predicts the next tokens to complete the suffix $s$.
Specifically, the model samples $\hat{x}_{i} \sim f_{\theta}(x_i \mid x_1, \ldots, x_{i-1})$ and subsequently feeds $\hat{x}_{i}$ back into the model to sample $\hat{x}_{i+1} \sim f_{\theta}(x_i \mid x_1, \ldots, \hat{x}_{i})$, iteratively.
Each newly generated token is conditioned on both the initial prompt and all previously generated tokens.
This decoding process is repeated until a termination condition is met, \textit{e.g.}, generating a special token \texttt{\textless{}/s\textgreater{}}, indicating the end of the sequence, or reaching a predefined maximum length of the token sequence.

\subsection{Memorization in Language Models}

Memorization, often seen as the antithesis of generalization, arises from overfitting, causing models to retain specific details of their training data~\cite{AlKaswan2024code_memorization_traces, Feldman2020_deep_long_tail}.
This phenomenon raises remarkable privacy concerns in the context of LMs, as these models may inadvertently memorize sensitive information and regurgitate it verbatim in response to certain prompts.

Extensive research has been undertaken to qualitatively and quantitatively examine memorization in LMs~\cite{Carlini2019secret_sharer, Carlini2021extracting, Carlini2023quantifying_memorization, AlKaswan2024code_memorization_traces, Huynh2023starcoder_memorization, Yang2024code_model_memorization, Niu2023codexleaks, Huang2024code_secret_memorization, Nie2024secret_memorization}.
Following these prior studies, we define memorization in LMs grounded in the extractability of training data.
In particular, we conceptualize memorization as a model's ability to store and reproduce exact pieces of information encountered during training.

\begin{definition}[Verbatim Memorization]
A string $s$ is considered memorized by an LM $f_{\theta}$ if there exists a prefix $p$ such that:
\begin{equation}
s = \argmax_{\hat{s}} f_{\theta}(\hat{s} \mid p) ~\land~ [p \mid\mid s] \in \mathcal{D} \,.
\label{equ_memorization}
\end{equation}
Here, $f_{\theta}(\hat{s} \mid p)$ denotes the model's likelihood of generating an entire sequence $\hat{s}$ given the prefix $p$, $[p \mid\mid s]$ represents the concatenation of strings $p$ and $s$, and $\mathcal{D}$ denotes the training dataset of $f_{\theta}$.
The $\argmax$ operation can be replaced by an appropriate decoding strategy (\textit{e.g.}, greedy sampling) to determine the model's outputs in practical applications.
\end{definition}

\noindent\textsc{\underline{Example 1.}}
Assume that the LM's training dataset contains a sequence ``\textit{\# Copyright (C) [2003] Daniel \textless{}daniel@gmail.com\textgreater{}}''. 
If the model is prompted with ``\textit{\# Copyright (C) [2003] Daniel \underline{\ \ \ \ \ }}'' and the most likely continuation is ``\textit{\textless{}daniel@gmail.com\textgreater{}}'', then the generated string is deemed memorized.

\section{Preliminary Study}
We first conduct a preliminary study to quantitatively examine the presence and severity of sensitive memorization in CLMs.

\subsection{Study Subjects}

\mypara{Studied CLMs}
To systematically analyze model memorization, we select representative CLMs varying in size and architecture. 
Following~\cite{AlKaswan2024code_memorization_traces, Yang2024code_model_memorization}, we examine four widely used models: CodeParrot-small (110M), CodeParrot (1.5B)~\cite{HuggingFace2022codeparrot}, and CodeGen-\{350M, 2B\}-Mono\cite{Nijkamp2023codegen}.
Our analysis also includes Qwen2.5-Coder-7B~\cite{hui2024qwen25coder}, a state-of-the-art CLM with over 35.3k monthly downloads on HuggingFace at the time of writing.
All selected models are accessible on HuggingFace Hub, enabling ethical and reproducible memorization analysis.

\mypara{Studied Datasets} 
We utilize \texttt{codeparrot-clean-train}~\cite{CodeParrot2022codeparrotclean}, a 50GB dataset comprising $\sim$5 million Python files.
We select it for two key reasons:
(1)~It is a high-quality, cleaned subset of GitHub corpora, extracted via Google's BigQuery~\cite{Google2016BigQuery}, making it representative of standard CLM training data.
(2)~It offers the repository source for each code instance, enabling realistic unlearning simulations where users request the removal of sensitive data unknowingly included in their repositories. 
These make it ideal for standardized memorization analysis and unlearning evaluations across CLMs.

\subsection{Memorization Quantification}

\subsubsection{\revise{Memorization Metrics}}
\label{sec_quantifying_memorization}

To accurately assess whether and to what extent CLMs retain specific data, our analysis adopts two quantitative metrics: \textit{Memorization Accuracy} (MA)~\cite{Tirumala2022memorization_without_overfitting} and \textit{Extraction Likelihood} (EL)~\cite{Jang2023knowledge_unlearning}.
As illustrated in~\autoref{tab_memorization_metrics}, these metrics measure memorization by comparing the CLM's generation with the true continuation, at the token and $n$-gram levels, respectively.

\mypara{\textit{Memorization Accuracy} (MA)~\cite{Tirumala2022memorization_without_overfitting}}
Given a specific token sequence $\mathbf{x} = (x_1, x_2, \ldots, x_N)$, we let the CLM $f_\theta$ sequentially process this sequence from left to right and predict each token based on its preceding context.
Then, MA calculates the accuracy of these predictions by comparing the generated tokens with their corresponding actual tokens in the sequence $\mathbf{x}$:
\begin{equation}
\text{MA}(\mathbf{x}) = \frac {\sum_{i=2}^{N} \mathbb{1} \{ \argmax_{\hat{x}_{i}} f_{\theta} (\hat{x}_{i} \mid x_{<i}) = x_i \}} {N - 1} \,,
\label{equ_ma}
\end{equation}
where $\mathbb{1}$ is the indicator function that returns 1 if the condition within the braces is true (\textit{i.e.}, the CLM's most likely prediction $\hat{x}_{i}$ accurately matches the actual token $x_i$) and 0 otherwise, and the notion $x_{<i}$ denotes the sequence of all tokens before position $i$.

\smallskip
\noindent\textsc{\underline{Example 2.}}
As shown in columns 1-3 of~\autoref{tab_memorization_metrics}, given various prefixes $x_{<i}$, $\hat{x}_{i}$ matches $x_i$ with 6 times (marked in \colorbox{matchgreen}{green}), resulting in an MA score of 0.5455.
We can see that MA measures the proportion of tokens in a sequence that the CLM can recall exactly, reflecting its capacity to memorize and reproduce training data.

\mypara{\textit{Extraction Likelihood} (EL)~\cite{Jang2023knowledge_unlearning}}
\revise{Compared with MA computed at the token level, EL enables a stricter standard for quantifying memorization}
by assessing the extent to which the generated sequence $\hat{x}_{\ge i}$ matches the true continuation $x_{\ge i}$ at the $n$-gram level:
\begin{gather}
\text{EL}_n (\mathbf{x}) \! = \! \frac {\sum_{i=2}^{N} \! \textsc{Overlap}_n (\argmax_{\hat{x}_{\ge i}} \! f_\theta (\hat{x}_{\ge i} \! \mid \! x_{<i}), x_{\ge i})} {N - n} \,, \nonumber \\
\textsc{Overlap}_n(\mathbf{a}, \mathbf{b}) = \frac {\sum_{c \in ng(\mathbf{a})} \mathbb{1} \{ c \in  ng(\mathbf{b})\}} {| ng(\mathbf{a}) |} \,,
\label{equ_el}
\end{gather}
where $\textsc{Overlap}_n$ measures the overlap of $n$-grams between two sequences, $ng(\cdot)$ denotes the list of $n$-grams within a sequence.
Higher $n$ values represent stricter standards for memorization quantification, adhering to higher privacy requirements.
Following~\cite{Jang2023knowledge_unlearning}, our study chooses $n$ values of 3, 5, and 10.

\smallskip
\noindent\textsc{\underline{Example 3.}}
As shown in columns 1 and 4-6 of~\autoref{tab_memorization_metrics}, 
in the first row, given the prefix ``This'', the number of the $3$-gram matches between $\hat{x}_{\ge i}$ and $x_{\ge i}$ is 2 (marked in \colorbox{matchgreen}{green}), leading to an $\textsc{Overlap}_3$ score of 0.2222.
After iterating through all the prefixes $x_{<i}$, all the $\textsc{Overlap}_3$ values are averaged to obtain a final $\text{EL}_3$ score of 0.1091.

\subsubsection{\revise{Memorization Thresholds}}
\label{sec_forgetting_threshold}

Memorization in CLMs varies, ranging from rarely reproduced sequences to verbatim repetition easily exploitable by adversaries. 
Without clear boundaries between them, it is difficult to prioritize and address genuine privacy vulnerabilities. 
To this end, we empirically establish explicit memorization thresholds based on the metrics $\text{MA}$ and $\text{EL}_n$:
\begin{equation}
    \begin{gathered}
        T_{\text{MA}} = \frac{1}{| \mathcal{D}^\prime |} \sum_{\mathbf{x}^\prime \in \mathcal{D}^\prime} \text{MA} (\mathbf{x}^\prime)\,,~T_{\text{EL}_n} = \frac{1}{| \mathcal{D}^\prime |} \sum_{\mathbf{x}^\prime \in \mathcal{D}^\prime} \text{EL}_n (\mathbf{x}^\prime)\,,
    \end{gathered}
\label{equ_forgetting}
\end{equation}
where $\mathcal{D}^\prime$ denotes a dataset consisting entirely of samples \textbf{\textit{unseen}} during the CLM's training phase. 
Intuitively, a training sample with memorization scores below these thresholds, appearing as if the model had never seen it, indicates safe memorization and is thus resistant to extraction attacks.
Conversely, samples exceeding these thresholds indicate potential risks of exposure and leakage.
The resulting memorization thresholds for the studied CLMs are presented in~\autoref{tab_forgetting_threshold}.

\begin{table}[!t]
    \small
    \setlength{\tabcolsep}{6pt} 
    \centering
    \caption{Memorization thresholds for the studied CLMs.}
    \vspace{-0.5em}
    \begin{tabular}{l|cccc}
        \hline
        \textbf{CLM} & \makecell{\textbf{MA (\%)}}  & \makecell{$\textbf{EL}_{\textbf{3}}$ \textbf{(\%)}}  & \makecell{$\textbf{EL}_{\textbf{5}}$ \textbf{(\%)}}  & \makecell{$\textbf{EL}_{\textbf{10}}$ \textbf{(\%)}}  \\
        \hline
        \text{CodeParrot-small} & 45.57 & 17.66 & 10.82 & 5.49 \\
        \text{CodeParrot} & 46.34 & 16.56 & 10.17 & 5.14 \\
        \text{CodeGen-350M-Mono} & 48.79 & 18.24 & 11.03 & 5.92 \\
        \text{CodeGen-2B-Mono} & 53.61 & 19.32 & 11.71 & 6.28 \\
        \text{Qwen2.5-Coder-7B} & 40.99 & 15.65 & 12.45 & 8.82 \\
        \hline
    \end{tabular}
    \vspace{-0.5em}
    \label{tab_forgetting_threshold}
\end{table}

\mypara{Unseen Dataset}
For the Qwen2.5-Coder-7B model, we compile $\mathcal{D}^\prime$ from two popular evaluation datasets, \textit{i.e.}, HumanEval~\cite{Chen2021human_eval} and MBPP~\cite{austin2021mbpp}, which have been explicitly excluded from the CLM's training corpus via data decontamination~\cite{hui2024qwen25coder}.
For the CodeParrot and CodeGen-Mono families, we compile $\mathcal{D}^\prime$ using a data crawling tool provided by~\cite{Xu2022code_llm_evaluation}, collecting 10,000 high-quality de-duplicated code files from GitHub repositories.
Repository selection criteria include:
each must have at least 500 stars and be created after the release dates of the CLMs.
Moreover, to address potential concerns that files might be copied from older versions already exposed to the CLMs, we only collect code files written in programming languages absent from the CLMs' training, \textit{e.g.}, Ruby, PHP, Rust, and Lua.
This strategy ensures that $\mathcal{D}^\prime$ is high-quality and genuinely unseen by the studied CLMs.

\subsection{Sensitive Memorization Identification}
\label{sec_sensitive_memorization_detection}

Not all memorization poses privacy risks; for instance, retaining public code snippets is far less concerning than memorizing private keys. 
While several techniques have been developed to extract memorized contents from CLMs~\cite{AlKaswan2024code_memorization_traces, Huynh2023starcoder_memorization, Yang2024code_model_memorization}, they mainly focus on analyzing non-sensitive code memorization. 
Recent studies~\cite{Huang2024code_secret_memorization, Niu2023codexleaks} have highlighted privacy risks by eliciting sensitive information from CLMs using well-crafted prompts.
However, they only reveal small-scale, isolated examples of sensitive memorization, lacking a systematic analysis of the broader extent of sensitive data retained by CLMs.
Thus, we aim to address the question: \textit{To what extent do CLMs memorize sensitive information from their training data?}

\mypara{Sensitive Data Identification}
To comprehensively identify sensitive data within code (\textit{e.g.}, emails, IP addresses, and API/SSH keys), we employ \texttt{detect-secrets}~\cite{Yelp2024detect-secrets}, a widely used regular expression-based detection tool, to scan the entire \texttt{codeparrot-clean-train} dataset. 
After filtering out local IPs and emails containing ``example'', we find that 939,665 out of 5,300,000 training samples (approximately 18\%) contain sensitive information.

\begin{figure}[!t]
	\centering
	\includegraphics[width=0.98\linewidth]{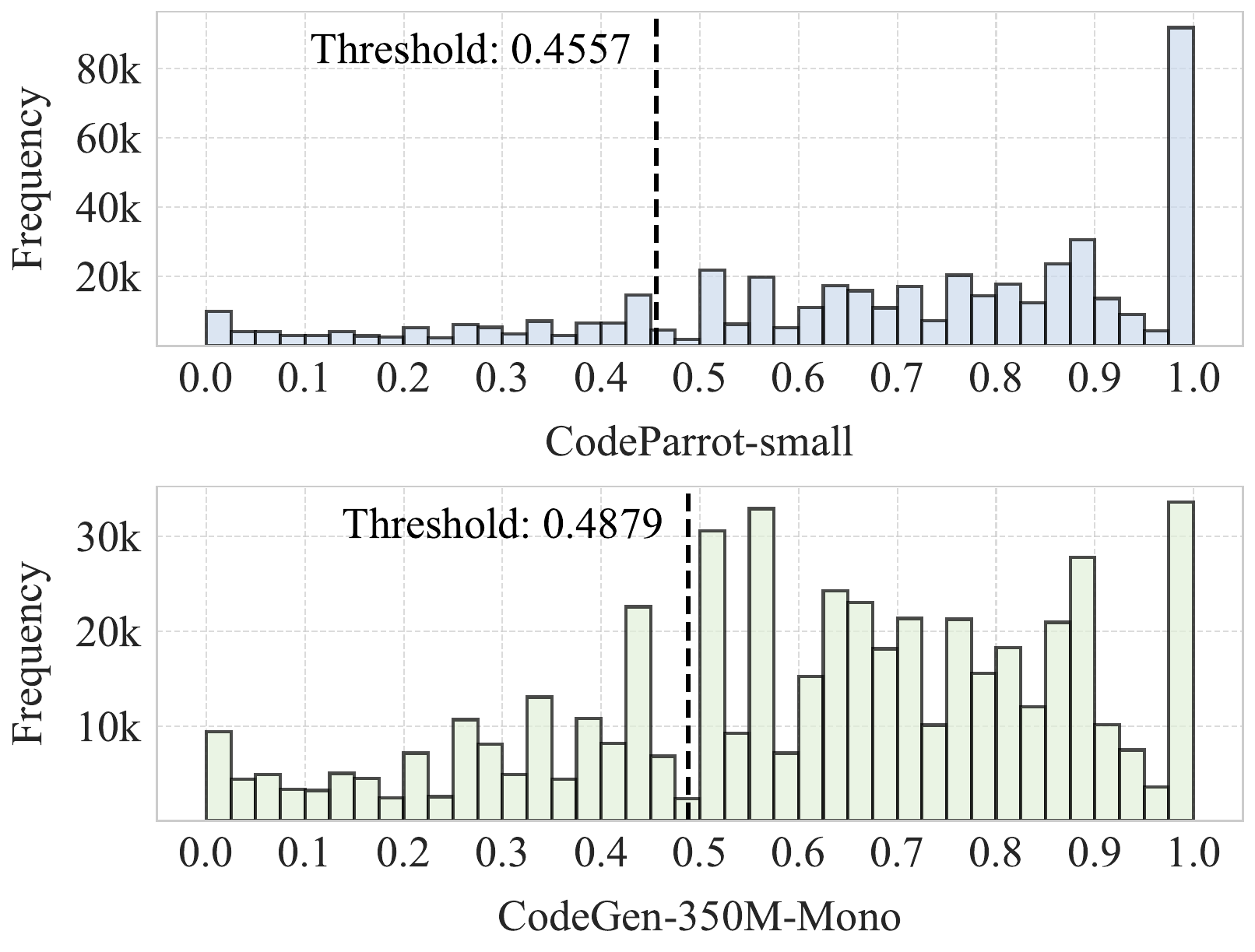}
        \vspace{-0.5em}
	\caption{The distribution of MA across sensitive data.}
	\label{fig_distribution}
    \vspace{-0.5em}
\end{figure}

\mypara{Sensitive Memorization Quantification}
We assess the memorization levels of sensitive segments in identified samples using the $\text{MA}$ metric.
$\text{MA}$ is preferred over $\text{EL}_n$ due to its efficiency in token matching, making it more suitable for large-scale analysis than the $n$-gram approach of $\text{EL}_n$.
Given computational constraints, we restrict our analysis to two relatively small models, \textit{i.e.}, CodeParrot-small and CodeGen-350M-Mono, and limit the examined samples to those containing sensitive data within a maximum token length (\textit{e.g.}, 512).
\revise{
Moreover, only sensitive segments are considered in this quantification; surrounding non-sensitive code is excluded.
For each sensitive segment, we prepend a fixed non-sensitive prefix (up to 128 tokens) when computing memorization. 
For instances containing multiple sensitive segments, we calculate the average $\text{MA}$ score across all segments.
These measures allow us to complete quantification on the full training dataset within 6 hours using a single GPU equipped with 80GB of memory.}

As shown in~\autoref{fig_distribution}, we find that 376,740 out of 473,994 training samples in CodeParrot-small and 363,806 out of 501,549 in CodeGen-350M-Mono (approximately 7\% of training data) exhibit sensitive memorization, with $\text{MA}$ scores exceeding the established memorization thresholds.

\begin{summary}
\textbf{Finding~\raisebox{-0.2ex}{\includegraphics[height=1em]{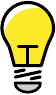}}:}
CLMs such as CodeParrot-small and CodeGen-350M-Mono memorize \textbf{approximately 7\%} training samples containing sensitive data, posing considerable privacy risks.
\end{summary}

\begin{figure*}[!t]
	\centering
	\includegraphics[width=0.98\textwidth]{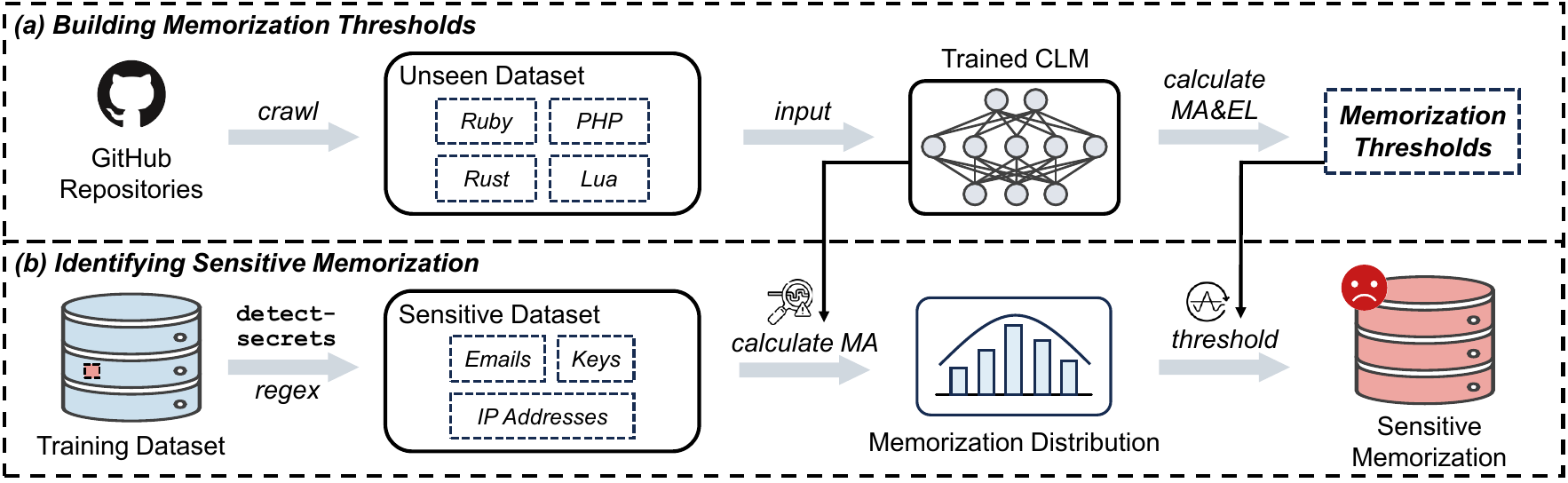}
        \vspace{-0.5em}
	\caption{An illustration of the sensitive memorization detection pipeline.}
        \vspace{-0.5em}
	\label{fig_memorization_detection_pipeline}
\end{figure*}

Building upon this finding, we extend our analysis to additional models, \textit{i.e.}, CodeParrot, CodeGen-2B-Mono, and Qwen2.5-Coder-7B.
For each studied CLM, we ultimately collect 10,000 highly memorized sensitive samples (\textit{e.g.}, $\text{MA} \ge 0.9$), resulting in \textbf{50,000} samples in total.
We compile them into a \textbf{Sensitive Memorization Dataset}, which documents the positions of all sensitive segments within each code sample along with their corresponding memorization scores. 
This dataset serves as the foundation for subsequent unlearning experiments, providing a standardized benchmark for evaluation.
The overall dataset collection pipeline is illustrated in \autoref{fig_memorization_detection_pipeline}.

\section{Unlearning Techniques}
\label{sec_unlearning}

Our preliminary study reveals that CLMs memorize substantial sensitive data from training corpora. 
To mitigate this issue, we explore unlearning techniques that enable targeted forgetting of memorized content.
We formally define the unlearning problem for CLMs and introduce three gradient ascent-based unlearning approaches: the vanilla method, the constraint-based method, and our proposed \systemnospace.

\subsection{Problem Statement}

Forgetting, the inverse of memorization, is typically studied in the context of  \textit{catastrophic forgetting}~\cite{Kemker2018measure_catastrophic_forgetting_in_nns, Kirkpatrick2017catastrophic_forgetting_in_nns}, where models lose prior knowledge when learning new tasks. 
These studies treat forgetting as an undesirable trait in training.
Recently, \citet{Jagielski2023measuring_forgetting} reinterpret forgetting positively, viewing it as a relaxed form of differential privacy.
However, they mainly examine passive forgetting during large-scale training.
In contrast, our study embraces an active form of forgetting, \textit{i.e.}, machine unlearning~\cite{Cao2015machine_unlearning}, which intentionally modifies trained models to erase previously memorized information.
Conceptually, we define forgetting as a reduction in the model's memorization of specific training samples.

Formally, let $f_\theta$ be a CLM trained on a dataset $\mathcal{D}$, and let $\mathcal{D}^f \! = \! \{\mathbf{x}^f\} \! \subset \! \mathcal{D}$ denote the \textit{forgotten set}, where each forgotten sample $\mathbf{x}^f \! = \! (x^f_1, \! x^f_2, \! \ldots, \! x^f_N)$ is a token sequence containing sensitive data. 
The set size $|\mathcal{D}^f| \! = \! k$ represents the number of samples undergoing unlearning simultaneously.
The goal of unlearning is to update the CLM to a new version, $f_\theta^\prime$, 
that no longer retains any information from each $\mathbf{x}^f$.
Specifically, after unlearning, each $\mathbf{x}^f$ should satisfy the following conditions, appearing as if never seen by the CLM:
\begin{equation}
    \begin{gathered}
        \text{MA} (\mathbf{x}^f) \le T_{\text{MA}}\,,~\text{EL}_n (\mathbf{x}^f) \le T_{\text{EL}_n}\,.
    \end{gathered}
\label{equ_forgetting}
\end{equation}

\subsection{Gradient Ascent-Based Unlearning}
\label{sec_gradient_ascent}

\begin{figure*}[!t]
	\centering
	\includegraphics[width=0.98\textwidth]{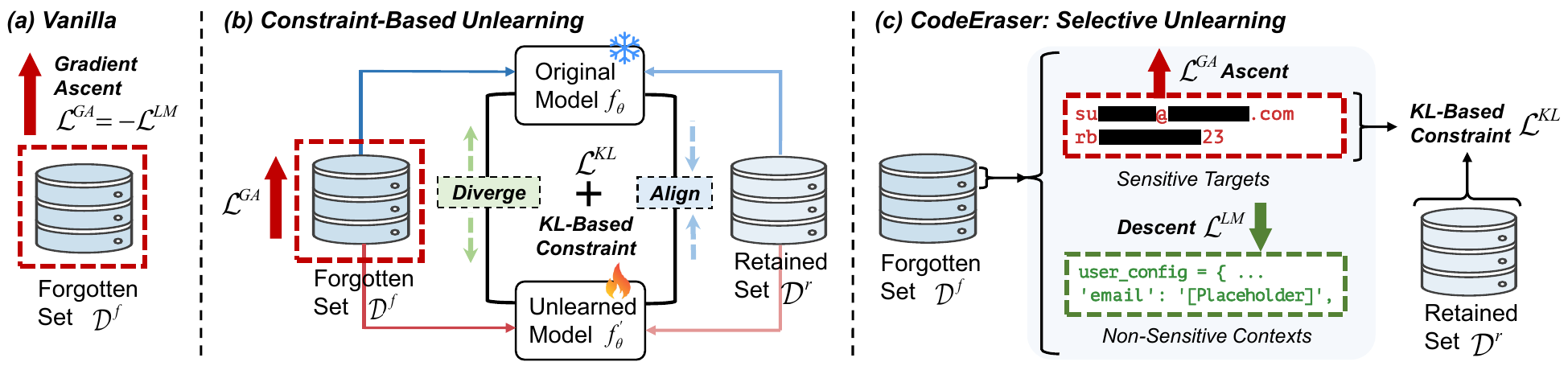}
        \vspace{-0.5em}
        \caption{
        \revise{An illustration of gradient ascent-based unlearning methods: (a) vanilla unlearning, (b) constraint-based unlearning, and (c) our proposed \systemnospace. Personal details are masked for ethical considerations.}
        }
	\label{fig_unlearning}
    \vspace{-0.5em}
\end{figure*}

\mypara{Vanilla Unlearning}
\textit{Gradient Ascent} (GA) ~\cite{Jang2023knowledge_unlearning, go2024each_unlearning, liu2022continual_private_unlearning} is a simple yet effective unlearning method designed to reduce the model's likelihood of predicting specific forgotten samples, thereby actively encouraging the removal of their information.
Specifically, for each $\mathbf{x}^f$, GA reverses the standard autoregressive language modeling objective by maximizing the negative log-likelihood, forcing the CLM to deviate from its original predictions.
Formally, as illustrated in~\autoref{fig_unlearning} (a), GA updates the unlearned CLM $f_\theta^\prime$ using the following loss function:
\begin{equation}
    \mathcal{L}^{GA} (\mathbf{x}^f) = -1 \cdot \mathcal{L}^{LM} (\mathbf{x}^f) = \log \prod_{i=1}^N f_{\theta}^\prime(x_i^f \mid x_1^f, \ldots, x_{i-1}^f) \,.
\label{equ_ga_loss}
\end{equation}

\mypara{Constraint-Based Unlearning}
A key challenge in unlearning is to remove targeted data without degrading the model's original utility.
Directly applying gradient ascent to the CLM may risk erasing unrelated yet valuable code knowledge. 
To address this, the \textit{Constraint-Based Unlearning} (CU) method~\cite{chen2023efficient_unlearn, yao2024llm_unlearning, Kurmanji2023unbounded_unlearn, yao2024machine_unlearning} seeks to minimize the \textit{Kullback-Leibler} (KL) divergence between the predictions of the original CLM $f_\theta$ and the unlearned CLM $f_\theta^\prime$ on the data to be retained, while maximizing divergence for the data targeted for forgetting.
Formally, given a \textit{retained set} $\mathcal{D}^r \! = \! \{(\mathbf{x}^r)\} \! \subset \! \mathcal{D}$, the CLM is updated using the following contrastive loss:
\begin{equation}
    \mathcal{L}^{K\!L} \! (\mathbf{x}^f\!, \! \mathbf{x}^r) \! = \! - \! \sum_{\mathbf{x}^f} \! K\!L\!(f_{\theta}\!(\mathbf{x}^f\!) || f_{\theta}^\prime\!(\mathbf{x}^f\!)) \! +  \alpha \cdot \! \sum_{\mathbf{x}^r}\! K\!L\!(f_{\theta}\!(\mathbf{x}^r\!) || f_{\theta}^\prime \!(\mathbf{x}^r\!)) \,,
\label{equ_kl_loss}
\end{equation}
where $\alpha$ is a hyperparameter controlling the balance between forgetting $\mathbf{x}^f$ and retaining $\mathbf{x}^r$.
Minimizing KL divergence on retained data ensures alignment with the original predictions, while maximizing it on forgotten data actively encourages effective forgetting.
In practice, as illustrated in~\autoref{fig_unlearning} (b), this KL divergence-based loss $\mathcal{L}^{KL}$ is typically combined with the GA-based loss $\mathcal{L}^{GA}$ for collaborative optimization:
\begin{equation}
    \mathcal{L}^{CU} = \mathcal{L}^{GA} (\mathbf{x}^f) + \lambda \cdot \mathcal{L}^{KL} (\mathbf{x}^f, \! \mathbf{x}^r)\,,
    \label{equ_cu_loss}
\end{equation}
where the hyperparameter $\lambda$ balances the intensity between gradient ascent updates and KL divergence-based constraints.

\mypara{\systemnospace: Proposed Selective Unlearning}
While techniques like gradient ascent and KL divergence-based constraint enable effective unlearning, they typically indiscriminately forget entire code samples, unnecessarily removing non-sensitive content. 
Motivated by insights from our preliminary study, we propose \systemnospace, an adapted unlearning method that selectively targets and erases sensitive memorized segments (\textit{e.g.}, API keys) without compromising the integrity and functionality of surrounding code.
\revise{
This segmentation design builds on the tool-based identification of sensitive elements within code (\textit{e.g.}, via \texttt{detect-secrets}) described in~\autoref{sec_sensitive_memorization_detection}, enabling accurate and targeted unlearning.}

Formally, for each forgotten sample $\mathbf{x}^f  \! = \! (x_1^f,  \! x_2^f,  \! \ldots, \! x_N^f) \! \in \! \mathcal{D}^f$, we segment sensitive sequences $\mathbf{s}^f  \! =  \! (s^f_1,  \! s^f_2,  \! \ldots,  \! s^f_m)$ from their non-sensitive contexts $\mathbf{c}^f  \! =  \! (c^f_1,  \! c^f_2,  \! \ldots,  \! c^f_n)$.
To achieve selective unlearning, we apply gradient \textbf{\textit{ascent}} exclusively on $\mathbf{s}^f$ to actively diminish their memorization, while applying gradient \textbf{\textit{descent}} on $\mathbf{c}^f$ to preserve their integrity. 
Accordingly, we apply a targeted KL divergence-based constraint on sensitive segments $\mathbf{s}^f$ rather than the entire sample $\mathbf{x}^f$.
Specifically, as illustrated in~\autoref{fig_unlearning} (c), we define the selective unlearning loss as follows:
\begin{equation}
    \mathcal{L}^{SU} = (\mathcal{L}^{GA}(\mathbf{s}^f) + \gamma  \cdot \mathcal{L}^{LM} (\mathbf{c}^f)) + \lambda \cdot \mathcal{L}^{KL} (\mathbf{s}^f, \! \mathbf{x}^r)\,,
    \label{equ_su_loss}
\end{equation}
where the hyperparameter $\gamma$ balances the trade-off between forgetting $\mathbf{s}^f$ and preserving $\mathbf{c}^f$.
This selective framework precisely restricts ascent updates to sensitive segments, minimizing the impact of unlearning on the CLM’s broader utility.

\smallskip
\noindent\textsc{\underline{Example 4.}}
Given a piece of code snippet ``\textit{user\_config = \{`email': `daniel@gmail.com', `password': `ABC'\}}'', the sensitive segments $s^f$ are ``\textit{daniel@gmail.com}'' and ``\textit{ABC}'', while the non-sensitive contexts $c^f$ are ``\textit{user\_config = \{`email': `[placeholder]', `password': `[placeholder]'\}}''. 
This segmentation preserves the original sequential structure required by autoregressive CLMs.
\smallskip

To stabilize the \textit{max-min} optimization of $\mathcal{L}^{KL}$ during unlearning, we adopt an iterative training strategy.
Specifically, we alternate training epochs between the forgotten set $\mathcal{D}^f$ and the retained set $\mathcal{D}^r$.
This strategy ensures balanced training dynamics, preventing either term from excessively dominating the optimization process.

\subsection{Connections and Discussion}
Here, we establish connections between our selective unlearning framework and other methods in~\autoref{fig_unlearning}.
Specifically, when removing the segmentation between sensitive and non-sensitive parts and 
setting the hyperparameters $\gamma$ and $\lambda$ in Eq.~(\ref{equ_su_loss}) to 0,
the selective unlearning loss collapses directly into the standard gradient ascent loss defined in Eq.~(\ref{equ_ga_loss}). 
In this scenario, the CLM indiscriminately performs gradient ascent on entire code instances rather than selectively targeting sensitive segments.
Similarly, by removing segment-level targeting in Eq.~(\ref{equ_su_loss}) and applying the KL divergence-based constraint to entire forgotten samples $\mathbf{x}^f$, the selective unlearning loss reverts to the original constraint-based formulation in Eq.~(\ref{equ_cu_loss}). 
In this case, the constraint-based unlearning considers whole-sample consistency, without explicitly distinguishing between sensitive and non-sensitive segments.
\revise{
These derivations demonstrate that our framework subsumes prior unlearning methods while extending them to support fine-grained, code-specific erasure of sensitive memorized segments.}

\section{Experiments and Analysis}

To evaluate the performance of various unlearning techniques for CLMs, we investigate the following \textit{Research Questions} (RQs):

\begin{itemize}[leftmargin=4mm, itemsep=0.05mm]
    \item \textbf{RQ1: Effectiveness and Efficiency.}
        \textit{How do the unlearning methods perform in terms of removing targeted sensitive information from CLMs (effectiveness) with minimal computational resources (efficiency)?}
    \item \textbf{RQ2: Model Utility Post-Unlearning.} 
        \textit{How do the unlearning methods affect the original utility of CLMs, particularly the code generation performance on the HumanEval benchmark?}
    \item \textbf{RQ3: Analysis on Forgotten Data.} 
        \textit{
        \revise{How do the characteristics of forgotten data (\textit{e.g.}, the number of samples $k$, their occurrence frequency in training, and the types of sensitive segments) impact unlearning performance?}
        } 
    \item \textbf{RQ4: Impact of Hyperparameters.} 
        \textit{\revise{How do hyperparameter settings (\textit{e.g.}, learning rate, $\gamma$, $\alpha$, and $\lambda$) impact unlearning effects?}}
\end{itemize}

\subsection{Experimental Setup}
\label{sec_implementation}

\mypara{Forgotten Set}
Following~\cite{Jang2023knowledge_unlearning}, we build the forgotten set for each studied CLM by \textbf{randomly sampling $k$ instances} from the corresponding sensitive memorization dataset, which are then subjected to unlearning.
To reduce the potential bias of random selection, we report the average result from 5 independent runs for each unlearning routine.
Unless otherwise specified, we report experimental results with a default setting of $k=32$.
We examine various values of $k$ from $\{8, 16, 32, 64, 128, 256, 512\}$, as detailed in~\autoref{sec_forgotten_data_analysis}, to demonstrate \systemnospace's scalability in handling varying numbers of unlearning requests from users.

\mypara{Retained Set}
The retained set is built to include non-targeted, non-sensitive data, serving as a benchmark for measuring the CLM's memorization retention after the unlearning process.
We leverage a code benchmark~\cite{AlKaswan2024code_memorization_traces} that offers 1,000 non-sensitive samples from BigQuery~\cite{Google2016BigQuery}. 
These samples have been demonstrated to be memorized by various CLMs, such as CodeParrot~\cite{HuggingFace2022codeparrot}, CodeGen~\cite{Nijkamp2023codegen}, and InCoder~\cite{Fried2023incoder}, making them suitable for building the retained set in our experiments.  
Specifically, for each studied CLM, we extract its corresponding memorized data and \textbf{randomly sample an equivalent number of $k$ instances} to form the retained set.

\mypara{Implementation Details}
In our experiments, the maximum token lengths are set to 512 for the forgotten set and 128 for the unseen dataset and the retained set, with any excess truncated. 
These lengths are chosen based on computational constraints while ensuring sufficient data is available for analysis.
For computing $\text{MA}$ and $\text{EL}_n$ scores, we adopt a greedy sampling strategy. 
Following~\cite{Jang2023knowledge_unlearning}, we set the global batch size equal to $k$ during unlearning. 
\revise{When processing a group of $k$ instances, we average their $\text{MA}$ and $\text{EL}_n$ scores to empirically decide whether they have been forgotten.} 
The learning rate is fixed at 3e-6, selected through empirical testing from the range \{1e-5, 8e-6, 5e-6, 3e-6, 1e-6\}, and we maintain a constant learning rate schedule throughout unlearning. 
Dropout and weight decay rates are both set to 0 to avoid regularization that might interfere with the unlearning process. 
We select $\alpha=1.0$ from \{0.5, 0.8, 1.0, 1.2, 1.5\}, and $\gamma=0.5$ and $\lambda=0.1$ from \{0.1, 0.2, 0.3, 0.4, 0.5\} through a systematic grid search, with detailed hyperparameter analysis in~\autoref{sec_parameter_analysis}.

\begin{table}[!t]
\small
\setlength\tabcolsep{5pt}
\captionsetup{width=\linewidth}
\caption{Evaluation of unlearning effectiveness. 
All values are reported as percentages (with \% symbol omitted).}
\vspace{-0.5em}
\centering
\renewcommand{\arraystretch}{1.05} 
\begin{tabular}{ll|ccccc}
    \hline
    \textbf{CLM} & \textbf{Method} & \textbf{MA} & $\textbf{EL}_{\textbf{3}}$ & $\textbf{EL}_{\textbf{5}}$ & $\textbf{EL}_{\textbf{10}}$ & \textbf{Red.} \\
    \hline
    \multirow{5}{*}{\makecell[l]{\textit{\textbf{CodeParrot}}\\\textit{\textbf{-small}}} } & 
    Original & 99.74 & 98.55 & 98.12 & 97.97 & - \\
    & \underline{\textit{Threshold}} & \underline{\textit{45.57}} & \underline{\textit{17.66}} & \underline{\textit{10.82}} & \underline{\textit{5.49}} & - \\
    & GA & 30.71 & 10.17 & 7.07 & 4.18 & 86.85 \\
    & CU & 22.14 & 7.79 & 6.19 & 4.72 & 89.69 \\
    & \system & 18.69 & 6.69 & 5.78 & 5.18 & 90.82 \\
    \hline
    \multirow{5}{*}{\textbf{\textit{CodeParrot}}} & 
    Original & 99.69 & 98.90 & 98.36 & 97.62 & - \\
    & \underline{\textit{Threshold}} & \underline{\textit{46.34}} & \underline{\textit{16.56}} & \underline{\textit{10.17}} & \underline{\textit{5.14}} & - \\
    & GA & 27.53 & 6.33 & 4.21 & 3.47 & 89.54 \\
    & CU & 24.18 & 6.11 & 4.39 & 3.09 & 90.48 \\
    & \system & 15.22 & 6.36 & 5.40 & 4.65 & 92.01 \\
    \hline
    \multirow{5}{*}{\makecell[l]{\textit{\textbf{CodeGen}}\\\textit{\textbf{-350M-Mono}}}} & 
    Original & 99.25 & 97.14 & 96.39 & 95.93 & - \\
    & \underline{\textit{Threshold}} & \underline{\textit{48.79}} & \underline{\textit{18.24}} & \underline{\textit{11.03}} & \underline{\textit{5.92}} & - \\
    & GA & 25.53 & 8.45 & 6.98 & 4.95 & 88.29 \\
    & CU & 18.65 & 6.98 & 5.73 & 4.88 & 90.75 \\
    & \system & 45.13 & 11.44 & 7.05 & 3.46 & 82.96 \\
    \hline
    \multirow{5}{*}{\makecell[l]{\textit{\textbf{CodeGen}}\\\textit{\textbf{-2B-Mono}}}} & 
    Original & 99.89 & 99.79 & 99.76 & 99.70 & - \\
    & \underline{\textit{Threshold}} & \underline{\textit{53.61}} & \underline{\textit{19.32}} & \underline{\textit{11.71}} & \underline{\textit{6.28}} & - \\
    & GA & 17.95 & 6.80 & 5.52 & 4.83 & 91.21 \\
    & CU & 11.80 & 6.40 & 5.99 & 5.54 & 92.55 \\
    & \system & 31.66 & 10.01 & 7.73 & 6.05 & 86.11 \\
    \hline
    \multirow{5}{*}{\makecell[l]{\textit{\textbf{Qwen2.5}}\\\textit{\textbf{-Coder-7B}}}} & 
    Original & 96.26 & 84.71 & 81.07 & 75.15 & - \\
    & \underline{\textit{Threshold}} & \underline{\textit{40.99}} & \underline{\textit{15.65}} & \underline{\textit{12.45}} & \underline{\textit{8.82}} & - \\
    & GA & 24.15 & 14.23 & 10.49 & 8.24 & 83.55 \\
    & CU & 16.63 & 8.77 & 6.84 & 5.48 & 89.16 \\
    & \system & 8.49 & 4.93 & 3.99 & 3.68 & 93.89 \\
    \hline
\end{tabular}
\label{tab_unlearning_effectiveness_efficiency}
\vspace{-0.5em}
\end{table}
\subsection{RQ1: Effectiveness and Efficiency}
\label{sec_effectiveness_efficiency}

We assess the effectiveness and efficiency of various unlearning techniques when applied to the studied CLMs.
\revise{In our context, effectiveness refers to the capability of the unlearning approach to successfully erase specific sensitive information retained by the CLM, while efficiency refers to the computational costs required to achieve this unlearning.}

\begin{figure}[!t]
	\centering
	\includegraphics[width=\linewidth]{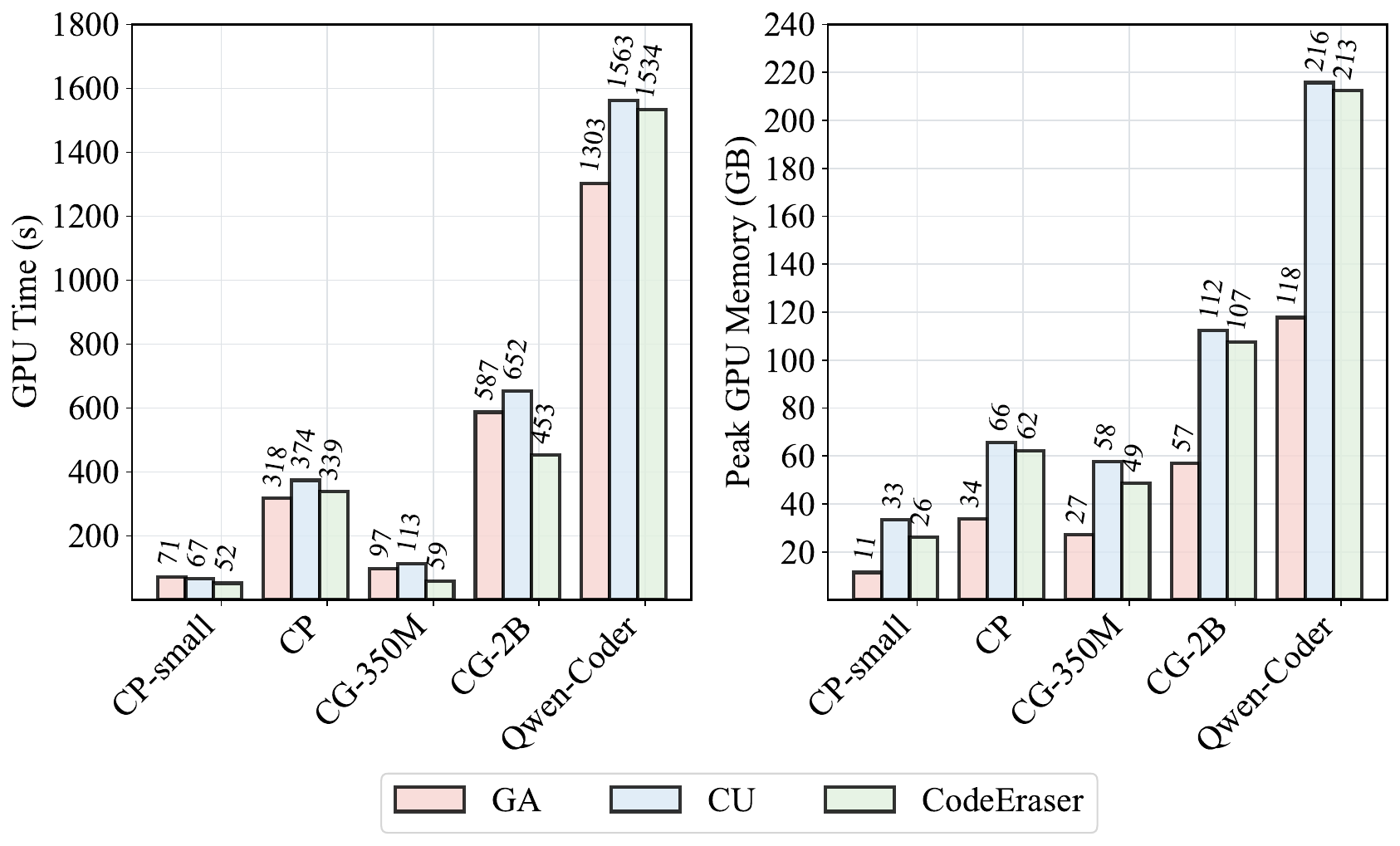}
        \vspace{-1.5em}
	\caption{Evaluation of unlearning efficiency.}
	\label{fig_efficiency}
    \vspace{-0.5em}
\end{figure}

\mypara{Unlearning Effectiveness}
To quantitatively evaluate the effectiveness of unlearning, we calculate MA, $\text{EL}_{3}$, $\text{EL}_{5}$, and $\text{EL}_{10}$ for the sensitive data targeted for removal in the forgotten set. 
An unlearning approach is deemed effective when the targeted sensitive data becomes difficult to extract from the model post-unlearning, characterized by MA and $\text{EL}_n$ scores falling below their respective \textbf{memorization thresholds} defined in~\autoref{sec_forgetting_threshold}. 
We also report the average memorization \textbf{reduction} rate of these metrics post-unlearning (abbreviated as \textit{Red.}).

As shown in~\autoref{tab_unlearning_effectiveness_efficiency}, the results indicate that \system achieves a substantial reduction in MA and $\text{EL}_{n}$ scores across all models, successfully lowering them below the predetermined memorization thresholds outlined in~\autoref{tab_forgetting_threshold}.
For instance, with the Qwen2.5-Coder-7B model, \system results in an average memorization reduction of \textbf{93.89\%}.
It is important to note that an unlearning method is considered effective as long as it reaches the forgetting criteria; it is not required to reduce more memorization than the baselines (\textit{i.e.}, GA and CU), as over-unlearning could lead to a loss of model utility.

\mypara{Unlearning Efficiency}
For efficiency, we measure the cumulative GPU time required to perform unlearning updates on the CLM until the memorization thresholds are reached for a group of $k=32$ instances. 
Additionally, we monitor peak memory usage across 4 GPUs during unlearning by leveraging PyTorch’s memory check API, and report the total footprint as the sum of these values.
These two metrics are chosen because they directly reflect the computational resources consumed during unlearning.

As shown in~\autoref{fig_efficiency}, with the Qwen2.5-Coder-7B model, our proposed \system completes the unlearning process within approximately 1500 seconds (averaging \textbf{46.88} seconds per sample), with a peak memory usage of around 200GB.
\revise{
This cost is considerably lower than alternatives such as differentially-private training or retraining the CLM after de-duplication, which are reported to typically require on the order of hundreds of A100 GPU days~\cite{Jang2023knowledge_unlearning}.}
Moreover, unlike the baselines that focus on the unlearning of entire code samples, \system exclusively targets the forgetting of specific sensitive data, enabling it to complete unlearning in a relatively shorter duration.
Although \system exhibits higher memory usage than GA (due to additional training steps on the retained set), it outperforms in terms of preserving the post-unlearning utility of CLMs, which will be further discussed in~\autoref{sec_model_utility}.

\begin{summary}
\textbf{Answer to RQ1:}
\revise{
\system demonstrates effectiveness and efficiency in erasing specific sensitive information from CLMs, thereby reducing potential security and privacy risks without incurring excessive computational costs.}
\end{summary}

\begin{table}[!t]
\small
\setlength{\tabcolsep}{5pt} 
\begin{minipage}[t]{\linewidth}
\captionsetup{width=\linewidth}
\caption{Evaluation of model utility post-unlearning. 
All values are reported as percentages (with \% symbol omitted). 
$\uparrow$ indicates that higher values correspond to better preservation of model utility. 
The best-performing unlearning method in each column is highlighted in \textbf{\underline{bold}}.}
\vspace{-0.5em}
\centering
\renewcommand{\arraystretch}{1.05} 
\begin{tabular}{ll|cccc}
    \hline
    \textbf{CLM} & \textbf{Method} & \textbf{P@1}$\uparrow$ & \textbf{P@5}$\uparrow$ & \textbf{P@10}$\uparrow$ & \textbf{Ret.}$\uparrow$ \\
    \hline
    \multirow{4}{*}{\makecell[l]{\textit{\textbf{CodeParrot}}\\\textit{\textbf{-small}}} }
    & Original & 3.48 & 4.56 & 4.96 & - \\
    & GA & 2.14 & 3.02 & 3.20 & 64.08 \\
    & CU & 2.62 & 3.43 & 3.66 & 74.77 \\
    & \system & \underline{\textbf{3.74}} & \textbf{\underline{4.59}} & \textbf{\underline{4.87}} & \textbf{\underline{102.10}} \\
    \hline
    \multirow{4}{*}{\makecell[l]{\textit{\textbf{CodeParrot}}} }
    & Original & 4.34 & 5.81 & 6.22 & - \\
    & GA & 2.08 & 3.29 & 3.83 & 55.38 \\
    & CU & 2.04 & 2.94 & 3.28 & 50.11 \\
    & \system & \textbf{\underline{3.86}} & \underline{\textbf{5.08}} & \underline{\textbf{5.63}} & \underline{\textbf{88.96}} \\
    \hline
    \multirow{4}{*}{\makecell[l]{\textit{\textbf{CodeGen}}\\\textit{\textbf{-350M-Mono}}} }
    & Original & 13.37 & 18.79 & 21.12 & - \\
    & GA & 11.68 & 16.59 & 18.51 & 87.76 \\
    & CU & 10.79 & 14.91 & 16.41 & 79.25 \\
    & \system & \underline{\textbf{13.36}} & \underline{\textbf{18.02}} & \underline{\textbf{19.96}} & \underline{\textbf{96.78}} \\
    \hline
    \multirow{4}{*}{\makecell[l]{\textit{\textbf{CodeGen}}\\\textit{\textbf{-2B-Mono}}} }
    & Original & 24.72 & 31.49 & 34.16 & - \\
    & GA & 21.20 & 28.34 & 31.40 & 89.23 \\
    & CU & 20.57 & 27.73 & 30.63 & 86.98 \\
    & \system & \underline{\textbf{23.00}} & \underline{\textbf{29.91}} & \underline{\textbf{32.94}} & \underline{\textbf{94.82}} \\
    \hline
    \multirow{4}{*}{\makecell[l]{\textit{\textbf{Qwen2.5}}\\\textit{\textbf{-Coder-7B}}}} & 
    Original & 61.07 & 73.61 & 77.23 & - \\
    & GA & 40.67 & 53.81 & 57.63 & 71.44 \\
    & CU & 48.54 & 64.70 & 69.59 & 85.83 \\
    & \system & \underline{\textbf{61.65}} & \underline{\textbf{73.41}} & \underline{\textbf{76.69}} & \underline{\textbf{99.99}} \\
    \hline
\end{tabular}
\label{tab_model_utility}
\vspace{-0.5em}
\end{minipage}
\end{table}

\subsection{RQ2: Model Utility Post-Unlearning}
\label{sec_model_utility}

\revise{
Ensuring robust privacy protections necessitates a delicate balance: 
the CLM must selectively forget targeted sensitive information to safeguard privacy without compromising its inherent capacity to perform general coding tasks.
We evaluate the efficacy of \system in achieving this balance.}

\mypara{Setup}
\revise{
To evaluate the impact of unlearning on the CLM's utility, we adopt the HumanEval benchmark~\cite{Chen2021human_eval}, a widely recognized standard for assessing code generation performance in CLMs~\cite{Bigcode2023benchmark}, with over 95.9k monthly downloads on HuggingFace at the time of writing.}
This benchmark measures the CLM's ability to solve programming tasks, where we report the Pass@1, Pass@5, and Pass@10 scores~\cite{Chen2021human_eval}, which measure the accuracy of generating correct solutions within 1, 5, and 10 attempts for each task, respectively (abbreviated as \textit{P@1}, \textit{P@5}, and \textit{P@10}).
By comparing these scores pre- and post-unlearning, we can observe the changes in the CLM's general coding performance.
To quantify these changes, we also report the average performance \textbf{retention} rate across these metrics post-unlearning (abbreviated as \textit{Ret.}).

\mypara{Results}
As shown in~\autoref{tab_model_utility}, \system has only a minor impact on model utility compared to the baselines.
Take Qwen2.5-Coder-7B as an example, \system preserves \textbf{99.99\%} of the CLM's code generation performance.
This lesser degree of degradation can be attributed to \systemnospace's sensitive information-targeted selective unlearning mechanism, which minimizes the impact of unlearning on model utility, ensuring that the code knowledge outside the specified forgotten set remains intact.

Among the baselines, a notable performance decline is observed in most cases when applying GA to the studied CLMs. 
This decline may stem from the gradient ascent updates, which, although performed only on the forgotten set, tend to \textit{soften} the probability distribution of generating each token across the vocabulary. 
This results in a more uniform distribution, which inadvertently dilutes the CLM's inherent knowledge base and reduces its overall utility.
Moreover, the CU approach does not demonstrate the expected level of utility preservation compared to GA in some cases. 
This may be due to the alignment of model behavior on shorter instances (128 tokens) being insufficient to offset the impact of forgetting longer instances (512 tokens). 
Instead, it could affect the stability of the model's updates, further compromising the integrity of the CLM's knowledge base. 
Given this unexpected phenomenon, we plan to investigate it further in future work to fully understand the dynamics of unlearning in CLMs.

\begin{summary}
\textbf{Answer to RQ2:}
\revise{\system has only a minor impact on the CLM's coding performance compared to baselines, validating the efficacy of our selective unlearning mechanism in preserving model utility while achieving targeted forgetting of sensitive information in code.}
\end{summary}

\subsection{RQ3: Analysis on Forgotten Data}
\label{sec_forgotten_data_analysis}

\revise{Recent studies have highlighted the importance of training data characteristics, such as duplication frequency and sensitive data type, in influencing memorization patterns of CLMs~\cite{AlKaswan2024code_memorization_traces, Carlini2023quantifying_memorization, Yang2024code_model_memorization}. 
These findings reveal the intricate nature of memorization in CLMs and imply a potential impact of such data attributes on the efficacy of unlearning.
To examine this, we evaluate \system on the CodeParrot model, focusing on targeted sensitive data samples that vary in number, duplication frequency, and type.}
For each setting, we report the average results of the HumanEval scores pre- and post-unlearning in 5 independent runs.

\begin{figure}[!t]
	\centering
	\includegraphics[width=\linewidth]{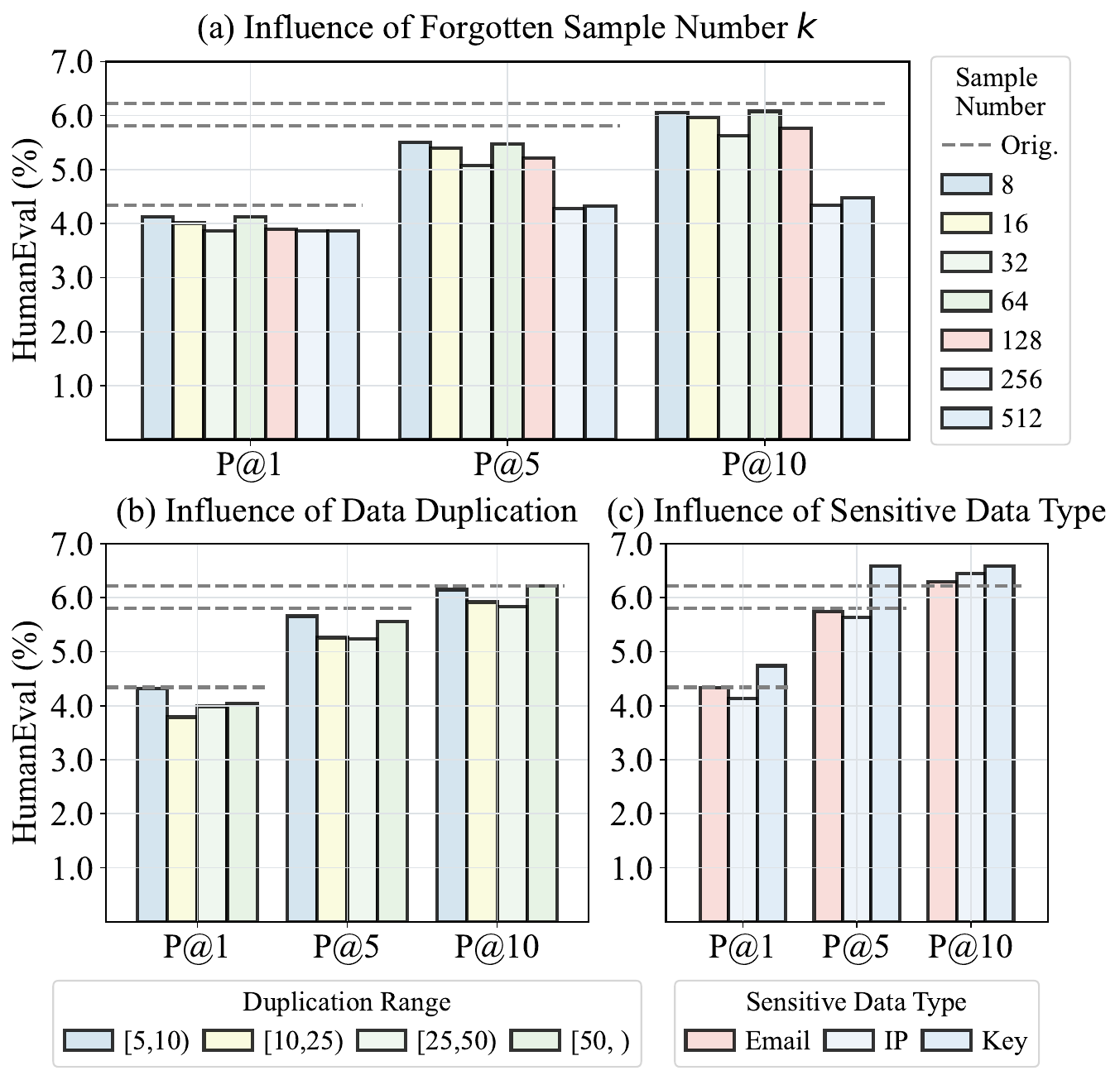}
        \vspace{-0.5em}
	\caption{Analysis on forgotten data. 
    Dashed lines ``- -'' represent the initial HumanEval scores of the CodeParrot model.}
	\label{fig_RQ3}
    \vspace{-0.5em}
\end{figure}

\mypara{Influence of Forgotten Sample Number $k$}
Our analysis examines how varying the number of forgotten samples ($k$) influences the efficacy of \systemnospace. 
\revise{As shown in~\autoref{fig_RQ3} (a), \system remains robust when unlearning a moderate number of sensitive samples (\textit{e.g.}, $k \le 128$), with the CLM effectively preserving its utility post-unlearning.
However, as the size of the forgotten set increases significantly (\textit{e.g.}, $k = 256$ or $k = 512$), utility scores such as P@5 and P@10 exhibit a noticeable decline. 
These results indicate potential scalability limitations in \systemnospace, particularly for larger-scale unlearning tasks involving extensive sensitive datasets (\textit{e.g.}, 10,000 samples). 
Ensuring effective unlearning at scale while minimizing utility degradation remains a key challenge, which we leave for future work.}

\mypara{Influence of Data Duplication}
To examine the impact of data duplications on unlearning, we utilize the duplication frequency statistics of training samples provided by the \texttt{codeparrot-clean-train} dataset. 
We roughly divide duplication frequencies into four ranges: $[5, 10)$, $[10, 25)$, $[25, 50)$, and $[50,~)$, which allows us to assess how varying levels of data duplication, from relatively low to very high frequencies, affect the unlearning process.
For each duplication level, we randomly select $k = 16$ samples that contain sensitive information to perform unlearning.

As shown in~\autoref{fig_RQ3} (b), the frequency of duplication significantly influences the CLM's utility post-unlearning. 
Interestingly, we can see that \system exhibits higher utility-preserving performance at the extremes of the duplication spectrum (\textit{i.e.}, $[5, 10)$ and $[50,~)$) compared to the intermediate duplication levels.
This phenomenon may be explained by the nature of the data involved. 
Low-duplicated memorized samples, potentially acting as outliers within the data distribution~\cite{Carlini2019secret_sharer}, may have a less entrenched influence on the model, making their removal less disruptive. 
On the other hand, highly duplicated samples are likely to cause model overfitting, meaning that their removal could reduce redundancy and mitigate overfitting, resulting in a minor impact on overall model performance.
These findings suggest that the impact of unlearning on model utility is not uniform across different levels of data duplication, and understanding these dynamics is crucial for optimizing our unlearning approach in the future.

\mypara{Influence of Sensitive Data Type}
To evaluate the influence of distinct sensitive data types (\textit{e.g.}, email, IP address, and API/SSH Key) on the unlearning process, we leverage the constructed sensitive memorization dataset for the CodeParrot model. 
For each type, we randomly select $k = 32$ samples containing only the corresponding sensitive data for unlearning.

As shown in~\autoref{fig_RQ3} (c), the influence on the CLM's utility post-unlearning varies depending on the type of sensitive data. 
This variation is likely due to differences in how the CLM memorizes these data types.
Surprisingly, the removal of API/SSH keys results in an improvement in model performance. 
This improvement may be attributed to the fact that, unlike emails and IP addresses, specific secret keys are usually atypical patterns within the data and are less likely to be heavily duplicated in the training dataset, making them outliers in the data distribution. 
Such atypical data outliers often distract the model and negatively impact its overall generalization.
Therefore, eliminating them can refine the CLM's representation space and shift its focus to more representative data, thereby enhancing overall performance.
Given these preliminary insights, future work will explore the effects of unlearning across a broader spectrum of sensitive data types.

\begin{summary}
\textbf{Answer to RQ3:}
The characteristics of the targeted sensitive data, \textit{e.g.}, number, duplication frequency, and type, significantly influence the CLM's utility post-unlearning. 
\revise{These findings reveal that unlearning effects are conditioned by data attributes, motivating future exploration of unlearning dynamics and robust strategies to minimize negative impacts on models.}
\end{summary}

\begin{figure}[!t]
	\centering
	\includegraphics[width=0.98\linewidth]{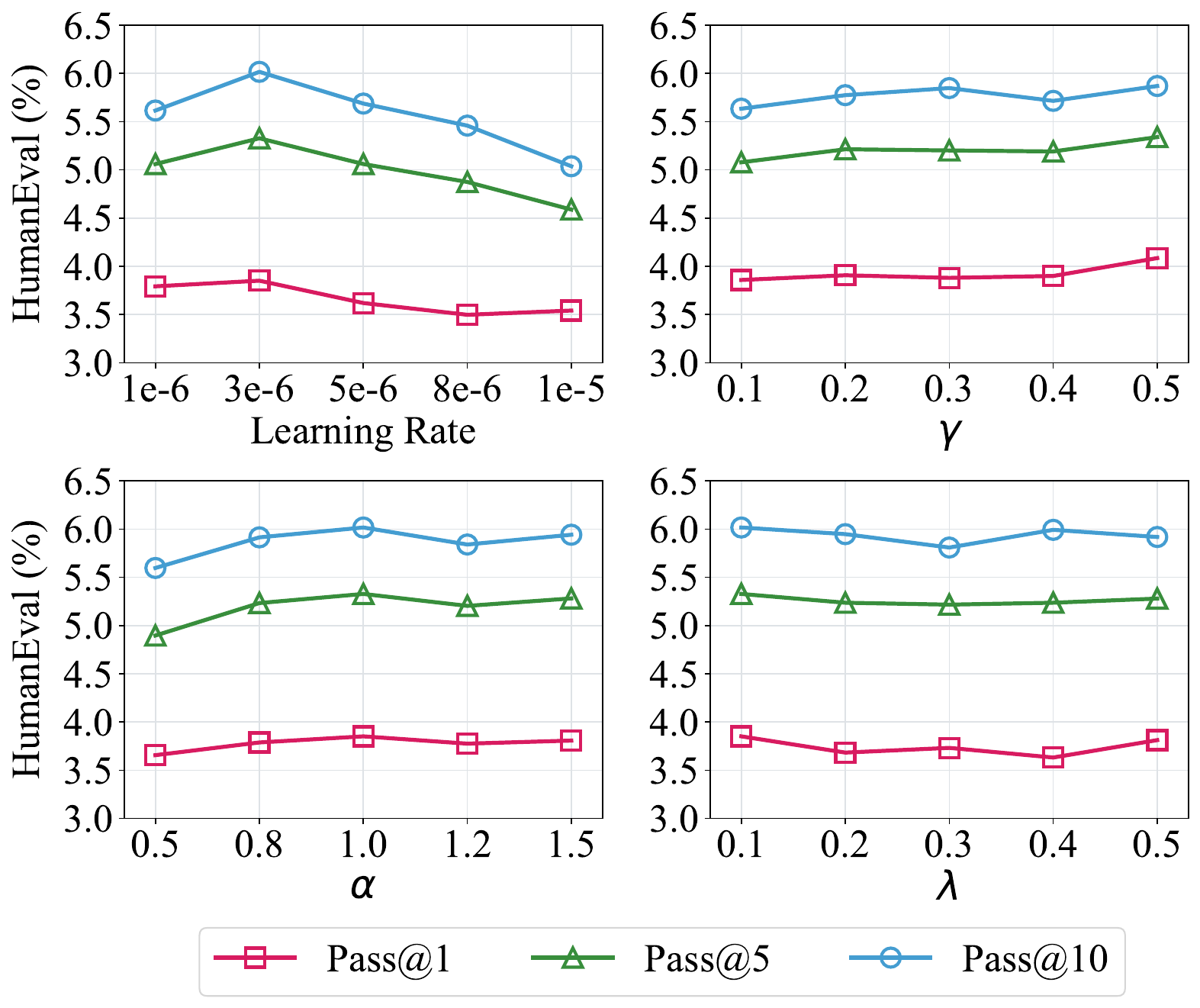}
        \vspace{-0.5em}
	\caption{Parameter analysis of learning rate, $\gamma$, $\alpha$, and $\lambda$.}
	\label{fig_parameter_analysis}
    \vspace{-0.5em}
\end{figure}

\subsection{RQ4: Impact of Hyperparameters}
\label{sec_parameter_analysis}

We analyze the impact of hyperparameters, including the learning rate and regularization factors (\textit{i.e.}, $\gamma$, $\alpha$, and $\lambda$), on model utility post-unlearning. 
As illustrated in~\autoref{fig_parameter_analysis}, the learning rate substantially influences model utility after unlearning, with excessively large values resulting in noticeable declines in utility metrics (\textit{e.g.}, P@5 and P@10). 
This underscores the importance of carefully tuning the learning rate to balance forgetting effectiveness against maintaining model performance.
In contrast, varying the hyperparameters $\gamma$, $\alpha$, and $\lambda$ within reasonable ranges results in only minor changes in post-unlearning utility, indicating robustness and flexibility of our method toward these parameters. Nonetheless, moderate values for these parameters are recommended, as overly aggressive settings may still negatively impact utility or insufficiently support effective forgetting.
Based on these insights, we select empirically determined optimal values for all hyperparameters to balance the trade-off between effective forgetting and model performance retention. The final settings employed in our experiments are detailed in~\autoref{sec_implementation}.

\begin{summary}
\textbf{Answer to RQ4:}
The learning rate substantially influences model utility after unlearning and must be carefully tuned. Regularization hyperparameters ($\gamma$, $\alpha$, $\lambda$) have comparatively minor impacts, allowing greater flexibility in their selection. \end{summary}

\section{Threats to Validity}

\mypara{Threats to Internal Validity}
\revise{Internal validity concerns whether our methodology introduces biases or errors that may distort the results. 
Our study identifies sensitive segments within code using a regular expression-based method and then quantifies their memorization.
However, this method constrains both the accuracy and coverage of secret detection, since regex rules can capture only a limited set of patterns.
To mitigate this threat, we employ \texttt{detect-secrets}~\cite{Yelp2024detect-secrets}, a state-of-the-art tool widely used for secret detection in large-scale code bases. 
This tool covers a broad spectrum of high-risk categories (\textit{e.g.}, API keys, tokens, and credentials) that are most relevant to real-world security incidents. 
Future work may incorporate additional detection methods~\cite{Basak2023secretbench, Huang2024code_secret_memorization, Feng2022secret_leakage_detection} to broaden coverage; however, such extensions are unlikely to alter our principal finding that CLMs manifest substantial sensitive memorization.}

\mypara{Threats to External Validity}
\revise{
External validity concerns the extent to which our findings can be generalized to other settings. 
This study focuses on three CLM families, \textit{i.e.}, CodeParrot, CodeGen, and Qwen2.5-Code, spanning 110M to 7B parameters; however, the results may not generalize to other CLMs.
To alleviate this threat, we select these families since they are widely adopted in research and practice, making them representative of the current CLM landscape. 
Moreover, the observed patterns are consistent across different model sizes within these families, suggesting that our findings are not tied to a specific scale. 
This limitation is also shared by many prior studies, which typically examine a few representative families rather than exhaustively covering all models. 
Thus, while additional CLM families might provide further evidence, our methodological choices sufficiently support external validity.}

\section{Related Work}
\mypara{Memorization in LMs}
Memorization, often seen as the antithesis of generalization, arises from overfitting, leading models to remember specific details of their training data~\cite{AlKaswan2024code_memorization_traces, Feldman2020_deep_long_tail}.
This phenomenon raises remarkable privacy concerns in the context of LMs, as these models may memorize and regurgitate sensitive information verbatim.
Extensive research has been undertaken to understand memorization in LMs qualitatively and quantitatively~\cite{Carlini2019secret_sharer, Carlini2021extracting, Carlini2023quantifying_memorization, Tirumala2022memorization_without_overfitting, Zhang2023counterfactual_memorization, Biderman2023predictable_memorization, Satvaty2024undesirable_memorization}.
Recent research~\cite{AlKaswan2024code_memorization_traces, Huynh2023starcoder_memorization, Yang2024code_model_memorization} has also explored memorization within CLMs, offering empirical studies to examine the extent to which CLMs inadvertently memorize and disclose their training data. 
Additionally, recent studies~\cite{Huang2024code_secret_memorization, Niu2023codexleaks} have highlighted privacy risks by extracting sensitive information from CLMs using well-crafted prompts.
Following this line, in this paper, we conduct a pioneering investigation into mitigating sensitive memorization in CLMs through machine unlearning.

\mypara{Machine Unlearning}
Machine unlearning, first proposed by Cao et al.~\cite{Cao2015machine_unlearning}, also known as \textit{selective forgetting}~\cite{Golatkar2020deep_network_scrubbing} or \textit{data removal/deletion}~\cite{Ginart2019data_deletion, Guo2020certified_data_removal}, aims to remove the influence of a specific set of training samples from the trained model.
Existing studies in this field can be categorized into two groups: 
\textit{1) Exact Unlearning: }
Exact unlearning seeks to remove specific samples' influence from the model completely. 
A straightforward method is to retrain the whole model from scratch after removing targeted data from the training set.
However, this method is computationally infeasible for large datasets.
Despite efforts to reduce the computational cost, they either primarily cater to simple machine learning models~\cite{Cao2015machine_unlearning, Ginart2019data_deletion} or rely on training data partitioning~\cite{Bourtoule2021sisa_unlearning, Chen2022graph_unlearning}, limiting their applicability to complex and large CLMs.
\textit{2) Approximate Unlearning:}
Approximate unlearning has recently emerged as a promising alternative, prioritizing efficiency and cost by relaxing the requirement for exactness.
Existing methods typically adjust the model's weights via gradient-based updates to approximate the weights of the model retrained from scratch~\cite{Golatkar2020deep_network_scrubbing, Guo2020certified_data_removal, Jang2023knowledge_unlearning, eldan2023whosharrypotterapproximate}. 
\revise{
Building on this paradigm, gradient ascent-based methods~\cite{Jang2023knowledge_unlearning, chen2023efficient_unlearn, yao2024machine_unlearning} have emerged as a dominant direction for efficient unlearning by reversing the learning of specific data, which also constitutes the focus of this study.
However, they often indiscriminately erase entire text instances rather than selectively targeting specific sensitive data (\textit{e.g.}, API keys embedded in code). 
While \citet{Wang2024selective_forgetting} proposed a heuristic that designates high-perplexity tokens in plain text as privacy tokens for unlearning, this approach is unsuitable for source code: 
identifiers are often assigned unique names with high perplexity, resulting in erroneous removal, whereas actual secrets such as API key strings typically follow predictable patterns with lower perplexity and may therefore escape removal.
In contrast, our approach employs a specialized tool (\textit{i.e.}, detect-secrets) to precisely identify secrets in code, enabling targeted unlearning while preserving the integrity and functionality of the surrounding code.}

\section{Conclusion}
In this paper, we pioneer the use of machine unlearning to erase sensitive memorization in CLMs. We first construct a novel dataset by systematically identifying and assessing high-risk code instances in the CLM’s training data. Then, we introduce \systemnospace, a selective unlearning approach that uses gradient ascent to remove sensitive information while preserving surrounding non-sensitive code via gradient descent. 
Additionally, \system employs a KL divergence-based constraint to maintain model utility post-unlearning. 
Extensive experiments on CodeParrot, CodeGen-Mono, and Qwen2.5-Coder demonstrate that \system effectively eliminates sensitive memorization while preserving overall model performance. 
Our study highlights the potential of unlearning in reinforcing data privacy in CLMs, providing a practical technique to actively mitigate the harms of model memorization.

\noindentparagraph{\textup{\textbf{Data Availability.}}}
All the experimental data and code used in this paper are available at \texttt{\url{https://github.com/CGCL-codes/naturalcc/tree/main/examples/code-unlearning}}.

\section*{Acknowledgment}
This work is supported by the Major Program (JD) of Hubei Province (Grant No. 2023BAA024).
We would like to thank all the anonymous reviewers for their insightful comments.

\balance
\bibliographystyle{ACM-Reference-Format}
\bibliography{ref}

\end{document}